\documentclass[conference]{IEEEtran}
\IEEEoverridecommandlockouts
\usepackage{cite}
\usepackage{amsmath,amssymb,amsfonts}
\usepackage{algorithmic}
\usepackage{graphicx}
\usepackage{textcomp}
\usepackage{xcolor}
\newcommand{\etal}{\textit{et al}. }
\def\BibTeX{{\rm B\kern-.05em{\sc i\kern-.025em b}\kern-.08em
    T\kern-.1667em\lower.7ex\hbox{E}\kern-.125emX}}
\begin{document}

\title{Self-Attention Based Multi-Scale Graph Auto-Encoder Network of 3D Meshes\\
\thanks{This project has received funding from the Normandy region under the COSURIA project. Part of this work was performed using the computing resources of CRIANN (Normandy, France).}
}

\author{
  \IEEEauthorblockN{Saqib Nazir\IEEEauthorrefmark{1}, Olivier Lézoray\IEEEauthorrefmark{1}, Sébastien Bougleux\IEEEauthorrefmark{1}} \\
  \IEEEauthorblockA{\IEEEauthorrefmark{1}\textit{Université Caen Normandie, ENSICAEN, CNRS, Normandie Univ,} \\
    \textit{GREYC UMR 6072, F-14000 Caen, France} \\
    \{saqib.nazir, olivier.lezoray, sebastien.bougleux\}@unicaen.fr
  }
}



\maketitle

\begin{abstract}
3D meshes are fundamental data representations for capturing complex geometric shapes in computer vision and graphics applications. While Convolutional Neural Networks (CNNs) have excelled in structured data like images, extending them to irregular 3D meshes is challenging due to the non-Euclidean nature of the data. Graph Convolutional Networks (GCNs) offer a solution by applying convolutions to graph-structured data, but many existing methods rely on isotropic filters or spectral decomposition, limiting their ability to capture both local and global mesh features. 
In this paper, we introduce 3D Geometric Mesh Network (3DGeoMeshNet), a novel GCN-based framework that uses anisotropic convolution layers to effectively learn both global and local features directly in the spatial domain. Unlike previous approaches that convert meshes into intermediate representations like voxel grids or point clouds, our method preserves the original polygonal mesh format throughout the reconstruction process, enabling more accurate shape reconstruction.  
Our architecture features a multi-scale encoder-decoder structure, where separate global and local pathways capture both large-scale geometric structures and fine-grained local details. Extensive experiments on the COMA dataset containing human faces demonstrate the efficiency of 3DGeoMeshNet in terms of reconstruction accuracy.

\end{abstract}


\section{Introduction}
3D meshes have become a crucial representation for capturing the geometric structure of complex objects, particularly in fields like computer vision, graphics, and animation. These meshes, composed of vertices, edges, and faces, provide an efficient and scalable way to represent surfaces, making them highly suitable for applications such as 3D shape reconstruction, shape modeling, face recognition, and shape segmentation \cite{review_recon_chen2023}. However, reconstructing 3D meshes presents unique challenges due to the non-Euclidean nature of the data, making the direct application of traditional neural network architectures, such as Convolutional Neural Networks (CNNs), difficult.

In Euclidean domains like images, CNNs have proven to be highly effective for tasks like image classification and natural language processing due to the structured nature of the data. However, 3D meshes are typically represented as graphs, where the number and orientation of neighbors for each vertex vary. This irregularity, combined with the non-Euclidean domain, complicates the use of standard convolutional architectures, which rely on consistent grid-like data structures. Early approaches, such as voxel-based representations and point clouds, were proposed to circumvent these challenges \cite{ren2023geoudf, ahmed2018survey}. However, they struggle to capture the fine geometric details of the meshes due to their high memory and computational costs.


Graph Convolutional Networks (GCNs) \cite{defferrard2016convolutional} have emerged as a promising solution to this problem, as they can operate directly on graph-structured data, enabling the learning of meaningful geometric features from 3D meshes. State-of-the-art (SOTA) GCN methods are typically classified into spectral and spatial approaches \cite{ranjan2018generating}. Spectral methods, such as ChebNet \cite{defferrard2016convolutional}, apply convolutions in the frequency domain but rely on isotropic filters, that are limited in their ability to capture both local and global features of complex shapes. Spatial methods, such as SpiralNet \cite{bouritsas2019spiral}, attempt to address this by defining a specific order of neighbors for each vertex, but these methods struggle with inconsistencies across different mesh topologies.


One of the earliest and most notable GCN-based approaches for 3D mesh reconstruction is the Convolutional Mesh Autoencoder (COMA) of Ranjan \etal \cite{ranjan2018generating}, which uses spectral convolution and mesh sampling to encode and decode 3D meshes. COMA utilizes spectral convolution layers and a quadric mesh up-and-down sampling \cite{qmc_garland1997surface} approach to encode and decode 3D meshes. The spectral approach transforms the mesh into the frequency domain using the Laplacian operator, which allows for more compact and efficient representations. While effective for aligned facial meshes, COMA struggles with larger-scale deformations and global variations, which are crucial for more complex human body reconstructions. 
To address this limitation, Bouritsas \etal \cite{bouritsas2019_spiral} introduced the Neural 3D Morphable Model (Neural3DMM), which replaced spectral convolution with a novel spiral convolution operator. 
In order to introduce anisotropic filters in graph convolutions, Bouritsas \etal \cite{bouritsas2019_spiral} formulated a spiral convolution operator (SpiralNet) that defines an explicit order of the neighbors via a spiral scan for each vertex on 3D meshes with fixed topology. However, serializing the local neighbors of vertices by following a spiral cannot resolve the inconsistency between different nodes. Furthermore, this method requires manually assigning a starting point to determine the order of neighbors, which makes it difficult to have the local coordinate system consistent across meshes. The selection of the starting point may affect the performance of the spiral convolution operator. Explicitly defining the order of neighbors cannot efficiently exploit the irregular structure of graphs.
Recent efforts, such as Local Structure-Aware Anisotropic Convolution (LSA-Conv), have sought to address these limitations by dynamically adjusting filter weights based on local geometry \cite{gao2022robust_LSA}. However, these methods often focus too heavily on local features, missing critical global topology, and requiring significant computational resources, especially for dense meshes.

In this paper, we propose 3DGeoMeshNet, a novel GCN-based framework designed to overcome the limitations of both spectral and spatial approaches. Our network uses anisotropic convolution layers that operate directly in the spatial domain, effectively capturing both global and local features without the need to convert the data into intermediate formats like voxel grids or point clouds. Our multi-scale encoder-decoder architecture features separate global and local pathways, allowing for efficient feature extraction and high-fidelity mesh reconstruction across varying scales.


The primary contributions of this work are:
\begin{itemize}
    \item We propose a 3D mesh reconstruction network that leverages an attention mechanism within a multi-scale autoencoder architecture, effectively capturing both local and global features.
    \item Our method operates entirely in the spatial domain, avoiding the complexity and limitations of spectral-domain methods, and directly processes 3D meshes without requiring conversion to intermediate representations.
    \item Our network achieves higher reconstruction accuracy compared to most of the isotropic filters or spectral decomposition-based approaches, offering a more efficient and scalable solution.
    \item Extensive experiments on the COMA \cite{ranjan2018generating} dataset, demonstrate the superiority of our method in terms of reconstruction accuracy.
\end{itemize}

\section{Related Work}
In this section, we review previous research efforts on 3D mesh reconstruction, focusing on both linear and non-linear methods. 3D representations of human bodies and faces are particularly challenging due to non-linear movements and varying expressions. Existing works are broadly categorized into linear statistical models and non-linear deep learning-based approaches.

\subsection{Linear Approaches}
The pioneering approach to 3D representation was introduced through 3D Morphable Models (3DMMs), which use statistical priors and dimensionality reduction techniques to model 3D faces, bodies, and hands. Blanz \etal \cite{blanz2023morphable} introduced the first linear parametric 3DMM based on Principal Component Analysis (PCA), which became the foundation for later models. The widely adopted Basel Face Model (BFM) \cite{paysan20093d} expanded on this work by incorporating a larger dataset of $200$ subjects, but it was still limited to neutral facial expressions. Li \etal \cite{FLAME_li2017} proposed FLAME to learn a unified model of facial shape and expression by analyzing 4D facial scan data, enabling the reconstruction and synthesis of dynamic facial expressions with high accuracy.

Later, Zhu \etal \cite{zhu2015high} merged the BFM with the FaceWarehouse dataset \cite{cao2013facewarehouse}, adding dynamic facial expressions and creating a more comprehensive model. For full-body reconstruction, the Skinned Multi-Person Linear Model (SMPL) \cite{loper2023smpl} became a prominent solution, using PCA to represent diverse body shapes in natural human poses. The MANO model \cite{romero2017embodied} applied similar principles for hand modeling, using a dataset of over $1,000$ high-resolution hand scans.



Recent advancements in linear methods, such as the Large-scale Facial Model (LSFM) introduced by Booth \etal \cite{booth2018large}, and Learning-by-Synthesis (LBS) \cite{li2017learning}, demonstrate the enduring relevance of these approaches. Despite the rise of non-linear methods, linear models are still valued for their simplicity, interpretability, and ability to capture global shape variations efficiently.


\subsection{Non Linear Approaches}

The rapid growth of deep learning applications on graph-structured data has led to significant advancements in non-linear methods for 3D mesh reconstruction. These approaches are particularly effective in capturing the complex and non-linear variations present in 3D meshes, such as human facial expressions and body deformations. The two dominant categories of Graph Neural Networks (GCNs) used for mesh reconstruction are spectral-based and spatial-based approaches.

Spectral-based methods define convolutional operations based on graph signal processing. One of the earliest models in this category is Spectral CNN \cite{rippel2015spectral}, which generalizes the concept of convolution to graphs using Laplacian eigenvectors. Building on this, ChebNet \cite{defferrard2016convolutional} and GCN \cite{kipf2016semi} significantly reduced the computational cost by using fast localized filters based on Chebyshev polynomials.
Yuan \etal \cite{yuan20193d} also used spectral graph convolutions to form a complete neural network architecture and introduced a new mesh pooling and de-pooling by repeated edge contraction, i.e., contracting two adjacent vertices to a new vertex. With the help of new pooling operations and convolutional kernels, they reported far fewer parameters than the original mesh Variational Auto-encoder (VAE) and thus can handle denser models. Jiang \etal \cite{jiang2019disentangled} propose an attribute decomposition framework for 3D face meshes that separates identity and expression components in a nonlinear manner using vertex-based deformation representation. 
Similarly, Zhang \etal \cite{zhang2020learning} introduces a variational autoencoder with a GCN, called Mesh-Encoder, to disentangle identity and expression components of 3D face shapes. 

These methods allow convolutions to be performed in the spectral domain, providing compact representations of the 3D mesh. However, a major limitation of spectral-based approaches is the need for eigen-decomposition, which is computationally expensive and difficult to scale for large meshes.

Spatial-based methods define graph convolutions based on the spatial relationships between nodes, making them more intuitive for 3D mesh reconstruction. One notable method is GraphSage \cite{hamilton2017inductive}, which samples a fixed number of neighbors for each node and aggregates neighboring features to compute new node representations. This method excels in capturing local geometry but may struggle with global feature extraction on large, complex meshes.
Graph Attention Networks (GAT) \cite{velivckovic2017_GATconv} introduced attention mechanisms that learn relative weights between connected nodes, improving the ability to capture varying levels of importance between neighbors. This approach allows for dynamic adjustment of weights, making it effective in handling non-uniform mesh topologies.
Monet \cite{palowitch2019monet} introduces pseudo-coordinates to define the relative spatial position between a node and its neighbors, allowing the network to assign different weights to neighboring nodes based on their positions. This flexibility makes MoNet particularly effective for 3D meshes where geometric features are irregularly distributed. A key contribution in the spatial-based category is FeaStNet \cite{verma2018feastnet}, which learns a dynamic weighting matrix for each convolutional operation based on node features. This allows FeaStNet to better handle irregular mesh structures by dynamically adjusting the influence of each neighboring node during the convolution process. 

Several mesh-based autoencoders have been developed to specifically address the reconstruction of 3D shapes. COMA \cite{ranjan2018generating}, one of the first autoencoder-based approaches for 3D meshes, used spectral convolutions for encoding and decoding facial meshes. While effective for small-scale facial reconstructions, COMA struggled with large deformations and global shape variations due to its reliance on spectral methods. 
SpiralNet \cite{bouritsas2019spiral} introduced a spiral convolution operator to capture local features of meshes by scanning neighboring vertices in a spiral order. Although SpiralNet improved upon COMA's local feature extraction, it faced challenges with mesh consistency and required manually defining the starting point for each spiral, which impacted the generalizability of the approach.
SpiralNet++ \cite{gong2019spiralnet++} addressed this problem by introducing a simple operator that captures local geometric structure from serializing the local neighborhood of vertices. Instead of randomly generating sequences per epoch \cite{bouritsas2019_spiral}, SpiralNet++ generates spiral sequences only once in order to employ the prior knowledge of fixed meshes, which improves robustness. 
Olivier \etal \cite{olivier2023facetunegan} presents FaceTuneGAN, a neural architecture adapting image-to-image translation networks for 3D face geometry. By leveraging SpiralNet++ \cite{gong2019spiralnet++} and large-scale face scan datasets, it achieves superior identity decomposition, facial expression transfer, and face neutralization. Sun \etal \cite{sun2022information} propose a VAE-based framework with dual decoders for disentangling identity and expression in 3D facial models. A mutual information regularizer enhances the balance between expressive and neutral face representations.
Local Structure-Aware Anisotropic Convolution (LSA-Conv) \cite{gao2022robust} built upon these approaches by dynamically learning anisotropic filters based on local geometry, similar to attention mechanisms. This allowed for better handling of irregular mesh data by adapting filter weights to capture local directional dependencies. However, LSA-Conv's focus on local geometry often missed global topology, which is critical for reconstructing large-scale 3D meshes. 

Unlike mesh-based methods, Gu \etal \cite{gu2023adversarial} used PointNet-based VAE and discriminator for reconstruction, disentangling 3D face identities and expressions.

\begin{figure*}[!t]
\centering
\includegraphics[width=\linewidth]{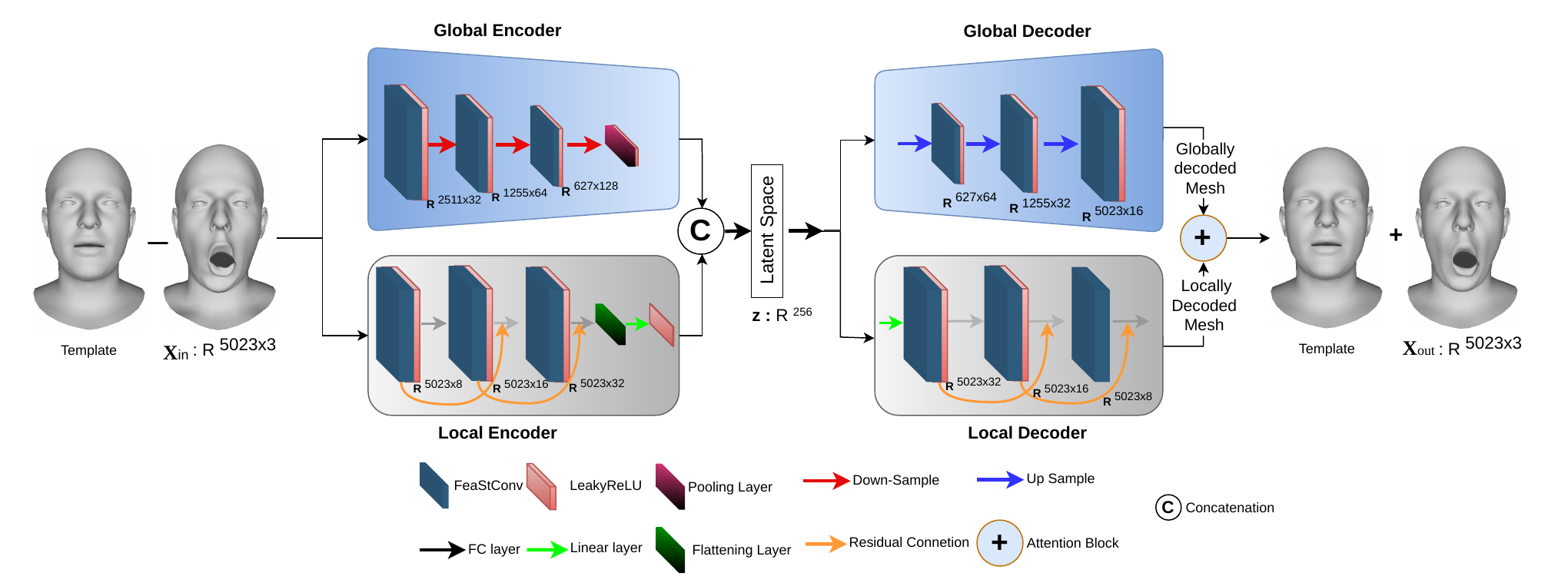}
\caption{The architecture of 3DGeoMeshNet used for training.}
\label{fig:architecture}
\end{figure*}

\section{Proposed Method}
\label{proposed_method}

\begin{figure}[!t]
\centering
\includegraphics[width=1\linewidth]{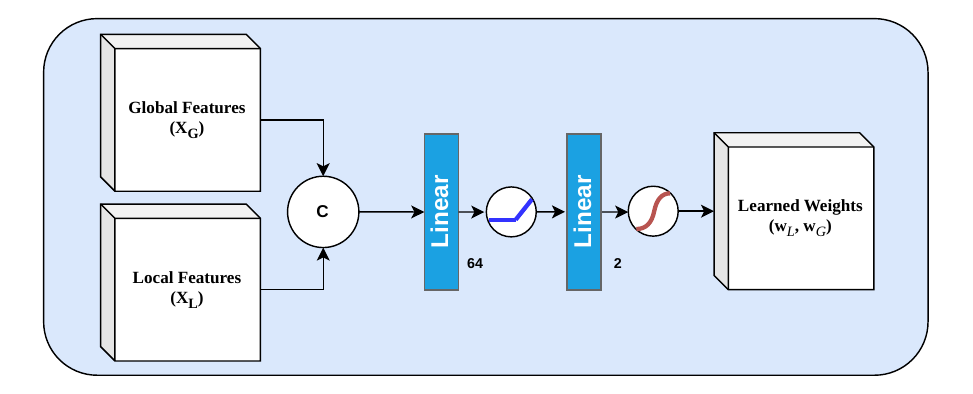}
\caption{Attention Block for features concatenation after the decoder.}
\label{fig:attention}
\end{figure}

We propose a framework called 3DGeoMeshNet, a GCN designed for 3D mesh reconstruction. The architecture of 3DGeoMeshNet is illustrated in Fig.~\ref{fig:architecture}. 
The architecture is based on an autoencoder structure that focuses on capturing local and global geometric features of the mesh. This is achieved through a combination of spatial GCs and hierarchical encoder-decoder networks.

\paragraph{Mesh representation}
The input of 3DGeoMeshNet is a 3D mesh $(V_\text{in},E_\text{in},F_\text{in},\mathbf{X}_\text{in})$, where $V$ is a set of $N$ vertices, $E$ is the set of edges, $F$ is the set of faces, and $\mathbf{X}\in\mathbb{R}^{N\times F}$ is a matrix of vertex features. Classically $F\,{=}\,3$ and $\mathbf{X}_\text{in}$ denotes the vertex positions. As for most other mesh autoencoders, 3DGeoMeshNet is trained to reconstruct meshes representing the same object type (\textit{e.\,g.}, scanned human heads), with the same number of vertices but with possibly different edges and faces. 
Moreover, the vertex set of each mesh is aligned with the vertex set of a template mesh $\mathcal{M}_\text{T}$, so that the input vertex positions are normalized as $\hat{\mathbf{X}}_\text{in}=(\mathbf{X}_\text{in}-\mathbf{X}_\text{T})/\sigma$, with $\mathbf{X}_\text{T}$ and $\sigma$ the mean and variance of the vertex positions on the training set.
As in \cite{hernandez2023deep}, to enhance the ability of the network to capture detailed geometric features, we add to $\hat{\mathbf{X}}_\text{in}$ the mean principal curvatures for each vertex (so $F=4$). It is defined as:

\begin{equation}
H(v)=\frac{1}{2}(k_1(v) + k_2(v))
\end{equation}

with $v\in V_\text{in}$, and \( k_1 \) and \( k_2 \) the estimated maximum and minimum principal curvatures of the surface at a given point \cite{meyer2003discrete}, respectively. As the mean principal curvature is invariant to rotation, it helps to understand the overall bending behavior of the surface at each vertex, as shown in \cite{shakibajahromi2024rimeshgnn}. 
\paragraph{Global architecture}
As illustrated in Fig. \ref{fig:architecture}, the autoencoder reconstructs vertex features from $\hat{\mathbf{X}}_\text{i}$ as $\hat{\mathbf{X}}_\text{out}=D(E(\hat{\mathbf{X}}_\text{in}))$, with $E:\mathbb{R}^{N\times F}\,{\rightarrow}\,\mathbb{R}^Z$ the encoder and $D:\mathbb{R}^{Z}\,{\rightarrow}\,\mathbb{R}^{N\times F}$ the decoder. The final mesh positions are $\mathbf{X}_\text{out}=\sigma^T\hat{\mathbf{X}}_\text{out}+\mathbf{X}_\text{T}$. As in FaceCom \cite{facecom_li2024}, the graph autoencoder merges a global encoder-decoder and a local encoder-decoder to capture large-scale features while maintaining fine-grained details. The encoders and decoders are GCNs based on Feature-Steered Graph Convolution (FeaStConv) \cite{verma2018feastnet}, which demonstrated excellent feature extraction capabilities compared to the other GC models \cite{cao2013facewarehouse, fey2019fast}. FeastConv aggregates features using the graph structure of the mesh equipped with learned dynamic weights on edges \cite{verma2018feastnet}. In the following, a GC layer is defined as FeastConv followed by LeakyReLU activation. The encoder $E$ and decoder $D$ are detailed below; the differences with FaceCom are discussed.

\paragraph{Encoder}
The encoder $E$ maps the input vertex features to a latent vector $\mathbf{z}=E(\hat{\mathbf{X}}_\text{in})\in\mathbb{R}^Z$ by concatenating global and local features extracted by a global encoder $E_\text{G}$ and a local encoder $E_\text{L}$, respectively, before applying a linear transformation with a Fully Connected (FC) layer:
\begin{equation}\label{eq:encoder_fusion}
    E(\hat{\mathbf{X}}_\text{in})={FC}(\texttt{concat}(E_\text{G}(\hat{\mathbf{X}}_\text{in}),E_\text{L}(\hat{\mathbf{X}}_\text{in})))
\end{equation}
Both encoders output a feature vector of size $Z/2$ before concatenation.
The \textit{global encoder} $D_\text{G}$ extracts features with a sequence of GC layers interleaved with down-sampling layers. Finally, the features are averaged across the vertices and rectified with LeakyReLU to produce a global feature vector. In contrast, the \textit{local encoder} $D_\text{L}$ extracts features with a sequence of GC layers with, contrary to \cite{facecom_li2024}, residual connections. Each residual connection skips a GC layer by linking its input directly to its output. This improves feature learning and mesh reconstruction, as shown in our ablation studies. The resulting vertex features are flattened and non-linearly transformed (Linear with LeakyReLU activation) into a local feature vector before concatenation with the global feature vector.
\paragraph{Decoder}
It generates vertex features  $\hat{\mathbf{X}}_\text{out}=D(\mathbf{z})$ from a latent code $\mathbf{z}$ by combining the features provided by the \textit{global decoder} $D_\text{G}$ and the \textit{local decoder} $D_\text{L}$:
\begin{equation}\label{eq:decoder_fusion}
    D(\mathbf{z})=\texttt{diag}(\mathbf{w}_\text{G})\,D_\text{G}(\mathbf{z}_\text{G})+\texttt{diag}(\mathbf{w}_\text{L})\,D_\text{L}(\mathbf{z}_\text{L})
\end{equation}
where $\left[\substack{\mathbf{z}_\text{G}\\\mathbf{z}_\text{L}}\right]={FC}(\mathbf{z})$ is the low-level feature vector decoded by an FC layer, split into two equal-length vectors ($\mathbf{z}_\text{G}$ and $\mathbf{z}_\text{L}$). The weight vectors $\mathbf{w}_\text{G},\mathbf{w}_\text{L}{\in}\,\mathbb{R}_{>0}^N$ control the contribution of each decoder at each vertex. 
Whereas these weights are empirically fixed scalars in FaceCom \cite{facecom_li2024}, we propose to adapt them to each vertex using an attention mechanism.
Symmetrically to the \textit{global encoder}, the \textit{global decoder} $D_\text{G}$ generates vertex features $\mathbf{X}_\text{G}=D_\text{G}(\mathbf{z}_\text{G})$ with a sequence of Up-Sampling and GC layers. 
Similarly, symmetrically to the \textit{local encoder}, the \textit{local decoder} $D_\text{L}$ generates vertex features $\mathbf{X}_\text{L}=D(\mathbf{z}_\text{L})$ 
by a Linear layer followed by three GC layers with residual connections.
\paragraph{Attention-based adaptive fusion}
As illustrated in Fig. \ref{fig:attention}, the weights $\mathbf{w}_G$ and $\mathbf{w}_L$ are adapted to the decoded vertex features $(\mathbf{X}_\text{G},\mathbf{X}_\text{L})$ using a small neural network:
\begin{equation}
\left[\mathbf{w}_\text{G},\mathbf{w}_\text{L}\right] = {Att}
(\mathbf{X}_\text{G},\mathbf{X}_\text{L})
\end{equation}
${Att}$ concatenates features for each vertex and defines attention weights through two linear layers with ReLU activation for the first layer and Softmax for the second. Combined with the decoder $D$, the resulting attention-based adaptive fusion allows the model to learn the optimal balance from the training set, rather than relying on a pre-fixed ratio as in FaceCom \cite{facecom_li2024}. Our ablation studies showed that the attention network improved model performance.
The attention mechanism is introduced in \cite{oktay2018attentionunetlearninglook} for U-Net architectures, and is used to focus the attention of the network on specific target structures and regions in the data and highlight only relevant information while training the model. The attention module has proven to improve the generalisability of the network. 
\paragraph{Down and up-sampling}
We considered the strategies presented in \cite{facecom_li2024},  improved from \cite{ranjan2018generating} based on Quadratic-Edge Collapse for mesh down-sampling. 
Each down-sampling operation reduces the number of vertices by a factor of $2$, and conversely each up-sampling operation doubles the number of vertices.


\section{Loss Function}

Our network is trained using a combination of two losses to ensure accurate mesh reconstruction while maintaining smoothness and structural coherence. 
During the training, the reconstruction loss is defined as the Mean Squared Loss (MSE).
\begin{equation}
    \text{L}_{\text{MSE}} = \frac{1}{N} ||\textbf{X}_{out}-\textbf{Y}||^2_{F}
\end{equation}

where $||\textbf{X}||_F=\sqrt{\sum_{i}\sum_{j}|X_{ij}|^2}$ is the Frobenius Norm, $\textbf{X}_{out}$ are the predicted features and $\textbf{Y}$ the ground truth ones.


To enforce a smooth latent space and maintain a spherical distribution for the latent representations, we use a custom spherical regularization loss. 

\begin{equation}
    \text{L}_{\text{reg}} = (||\textbf{z}||_2 -1)^2
\end{equation}

This ensures that the latent features $\textbf{z}$ lie close to a spherical manifold, promoting better generalization. This loss is particularly effective in preventing vanishing
KL-divergence and non-decreasing $\text{L}_{\text{MSE}}$ and maintaining structural integrity in the reconstructed meshes. Our experimental results further supported the notion that this approach facilitates optimization-based inference procedures. A previous study \cite{facecom_li2024} has illustrated that this approach can significantly enhance autoencoder generative capacity and overall stability. 

The overall reconstruction loss is defined as:
\begin{equation}
    \text{Total Loss} = \text{L}_{\text{MSE}} + \lambda_{\text{reg}} \cdot \text{L}_{\text{Reg}}
\end{equation}
where $\lambda_{\text{reg}}= 0.0001$ is a weighting factor for the spherical regularization loss. This combination ensures a balance between accurate reconstruction, robustness to outliers, and smooth latent representations.



\section{Experiments}
In this section, we first evaluate the proposed model on the COMA 3D shape dataset by comparing them to SOTA approaches for autoencoder-based reconstruction. In addition, ablation tests are conducted to demonstrate the effectiveness of our model. Finally, we show the results of two applications 3D interpolation/extrapolation and mesh denoising to demonstrate the effectiveness of our method for real-world problems.
\subsection{Datasets and Training Details}
We evaluate our model on the COMA \cite{ranjan2018generating} dataset. COMA is a human facial dataset that consists of $12$ classes of extreme expressions from $12$ different subjects. The dataset contains $20466$ 3D meshes that were registered to a common reference template with $N=5023$ vertices.
We split the COMA dataset into training and test sets with a ratio of $9:1$ and randomly selected $100$ samples from the training set for validation same as in \cite{ranjan2018generating}. 



We initialized the learning rate at $0.0005$ and halved it every $50$ epoch. The training process utilized the Adam optimizer within the PyTorch framework, with the entire training process taking approximately $24$ hours on an NVIDIA Geforce $RTX4090$ GPU. The batch size is $32$ and the total epoch number is $300$. 
The local convolutional layers are configured with feature sizes of 8, 16, and 32, while the global convolutional layers are designed with feature sizes of 32, 64, and 128.
For the quantitative results, we calculate the mean, median, and $L2$ errors in millimeters (mm).

\subsection{Results}
For the evaluation of 3D mesh reconstruction, we compared the performance of our proposed method, 3DGeoMeshNet, against several SOTA approaches, including PCA \cite{blanz2023morphable}, FLAM \cite{FLAME_li2017}, Jiang \etal \cite{jiang2019disentangled}, Zhang \etal \cite{zhang2020learning}, FaceTuneGAN \cite{olivier2023facetunegan}, Sun \etal \cite{sun2022information},
COMA \cite{ranjan2018generating}, SpiralNet++ \cite{gong2019spiralnet++}, Yuan \etal \cite{yuan20193d}, Gu \etal \cite{gu2023adversarial}, FaceCom \cite{facecom_li2024}, and LSA-Conv \cite{gao2022robust_LSA}. These methods vary significantly in their underlying architectures: PCA represents a linear model, while all other methods employ Deep Neural Networks (DNN) with non-linear mechanisms. Moreover, Gu \etal \cite{gu2023adversarial} utilizes PointNet, whereas the remaining methods adopt GCNs. FaceCom \cite{facecom_li2024} is designed for 3D in-painting but we have modified it and show its results for 3D mesh reconstruction as well.  

\begin{table}
    \centering
    \caption{Reconstruction Errors of PCA, FLAM, Jiang \etal, Zhang \etal, FaceTuneGAN, Sun \etal, COMA, SpiralNet, SpiralNet++, Yuan \etal, GU \etal, FaceCom, LSA-Conv and ours on COMA dataset.}
    \begin{tabular}{c|c|c|c}
        Method & \textbf{z} & Mean  & Median   \\ \hline
        PCA \cite{blanz2023morphable} & -- & 1.639 ± 1.638 & 1.101   \\
        FLAME \cite{FLAME_li2017} & -- &  1.451 ± 1.64 &  0.87   \\
        Jiang \etal \cite{jiang2019disentangled} & -- &  1.413 ± 1.639 &  1.017   \\
        COMA \cite{ranjan2018generating} & 8 & 0.845 ± 0.994  & 0.496  \\
        FaceTuneGAN \cite{olivier2023facetunegan} & -- &  0.83 ± 0.21 &  0.77   \\
        Zhang \etal \cite{zhang2020learning} & -- &  0.665 ± 0.748 &  0.434   \\
        Sun \etal \cite{sun2022information} & -- &  0.663 ± 0.215 & 0.643   \\
        Gu \etal \cite{gu2023adversarial} & 256 &  0.651 ± 0.208  &  0.625 \\
        Yuan \etal \cite{yuan20193d} & 128 & 0.583 ± 0.436  & -- \\
        SpiralNet++ \cite{gong2019spiralnet++} & -- & 0.54 ± 0.66 & 0.32 \\
        FaceCom (modified)\cite{facecom_li2024} & 256 &  0.516 ± 0.708  &  0.255  \\
        \textbf{LSA-Conv \cite{gao2022robust_LSA}} & 32 & \textbf{0.153} ± 0.217 & \textbf{0.077} \\
    \hline
        3DGeoMeshNet (Ours) & 256 & 0.171 ± \textbf{0.187} & 0.105 
    \end{tabular}
    \label{tab:sota_results_coma}
\end{table}

Table \ref{tab:sota_results_coma} presents the quantitative comparison of reconstruction errors for our method and related approaches on the COMA dataset. With a latent dimension of $\textbf{z}=256$, 3DGeoMeshNet achieves the lowest reconstruction error, outperforming all the competing methods by a significant margin except LSA-Conv. However, we believe that incorporating the LSA-Conv instead of FeastConv in our architecture can further enhance our model's results. 
Notably, DNN-based methods consistently outperform PCA, as linear models like PCA are inherently limited to capturing global features, while DNN-based non-linear models leverage convolutional operations to capture both global and local features. These results demonstrate the efficacy of incorporating global and local encoder-decoder networks, which enhance the expressive power of 3D shape representation compared to ChebNet-based COMA and SpiralNet.

LSA-Conv \cite{gao2022robust_LSA} utilizes a smaller latent dimension ($\textbf{z}=32$) and achieves competitive results. When comparing 3DGeoMeshNet to Yuan \etal \cite{yuan20193d} and Gu \etal \cite{gu2023adversarial}, with latent dimensions of $\textbf{z}=128$ and $\textbf{z}=256$, respectively, our method achieves superior reconstruction accuracy, particularly when using the same latent dimension as Gu \etal. 
In our experiments, FaceCom \cite{facecom_li2024} was used as the baseline model. The quantitative comparison with FaceCom demonstrates that our modified network significantly outperformed the baseline. Specifically, the modified FaceCom achieved a mean error of $0.516$, whereas our approach achieved a substantially lower mean error of $0.171$. These results highlight the effectiveness of incorporating the attention module, residual connections, and principal curvature into the network design. 

\begin{figure}[!t]
\centering
\includegraphics[width=1\linewidth]{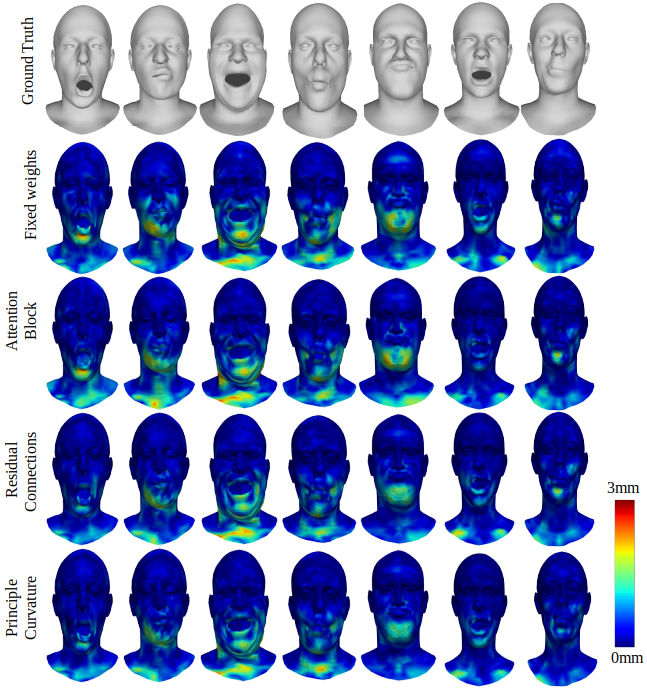}
\caption{Ablation results on COMA dataset.}
\label{fig:results_CMA}
\end{figure}

Fig. \ref{fig:results_CMA} illustrates qualitative results of the reconstruction errors for selected examples from the COMA test set. Using a latent size of $\textbf{z}=256$, our method consistently produces smaller reconstruction errors across various test cases, underscoring its robustness and accuracy.

\begin{table*}[!h]
    \centering
    \caption{Ablation studies on COMA dataset. At first, we increased the latent size from $32$ to $256$ and later added the Attention module, Residual connections, and Mean Principal Curvature. }
    \begin{tabular}{c|c|c|c|c}
        Method & Latent \textbf{z} & Mean Error & Median Error & L2  \\ \hline
        3DGeoMeshNet & 32 & 0.831 ± 0.964 & 0.489 & 0.735   \\ 
        3DGeoMeshNet & 64 & 0.632 ± 0.795  & 0.342  & 0.695  \\ 
        3DGeoMeshNet & 128 & 0.522 ± 0.721  & 0.259  &  0.514  \\ 
        3DGeoMeshNet & 256 & 0.516 ± 0.708  & 0.255  & 0.506  \\ \hline
        3DGeoMeshNet (Attn) & 256  & 0.223 ± 0.217 & 0.153 & 0.179   \\
       3DGeoMeshNet (Attn+Res) & 256 & 0.177 ± 0.191 & 0.110 & 0.150  \\
       \textbf{3DGeoMeshNet (Attn+Res+Mean)} & \textbf{256} & \textbf{0.171 ± 0.187} &  \textbf{0.105} & \textbf{0.146}  \\
        \hline
        3DGeoMeshNet (Local path only) & 256 &  0.688 ± 0.527 & 0.563 & 0.797  \\
        3DGeoMeshNet (Global path only) & 256 &  1.520 ± 0.705 &  1.425 & 0.973  \\

    \end{tabular}
    \label{tab:results_ablation}
\end{table*}

\subsection{Ablation Study}


Table \ref{tab:results_ablation} provides an ablation analysis of 3DGeoMeshNet under different configurations. Initially, we evaluated the model using a latent size of $\textbf{z}=32$ and progressively increased it to $\textbf{z}=64$, $\textbf{z}=128$, and $\textbf{z}=256$. As the latent dimension increased, the number of model parameters grew accordingly, leading to a reduction in reconstruction error. This trend illustrates the trade-off between model complexity and performance. Specifically, the mean reconstruction error decreased from $0.831$ to $0.516$ when the latent size increased from $\textbf{z} = 32$ to $\textbf{z} = 256$.  


To further enhance the model, we incorporated an attention module, residual connections, and principal curvature as additional input features. Adding the attention module reduced the mean reconstruction error from $0.516$ to $0.223$, while residual connections further improved it to $0.177$. Finally, incorporating the mean principal curvature as an additional input achieved the lowest mean reconstruction error of $0.171$.

To evaluate the significance of the dual-path architecture, we conducted an ablation study by removing either the global or local path and, consequently, the attention module from our model.
As shown in the final rows of Table \ref{tab:results_ablation}, utilizing only the local path resulted in a substantial performance decline, with the mean error increasing from $0.177$ to $0.688$. More notably, when employing only the global path, performance deteriorated even further, with a mean error of $1.520$. These results demonstrate the importance of integrating both global and local paths with an attention-based fusion mechanism. The global path appears to capture high-level structural information that complements the local geometric features processed by the local path, with the attention module effectively combining these complementary representations to achieve optimal reconstruction quality.

Fig. \ref{fig:results_CMA} illustrates qualitative insights into the impact of these modifications. Inclusion of the attention module, residual connections, and principal curvature significantly improves the reconstruction quality, as evidenced by the reduced errors in the reconstructed meshes. These results demonstrate the effectiveness of our architectural enhancements in achieving good performance for 3D mesh reconstruction.



\subsection{Applications}

\begin{figure}[!t]
\centering
\includegraphics[width=1\linewidth]{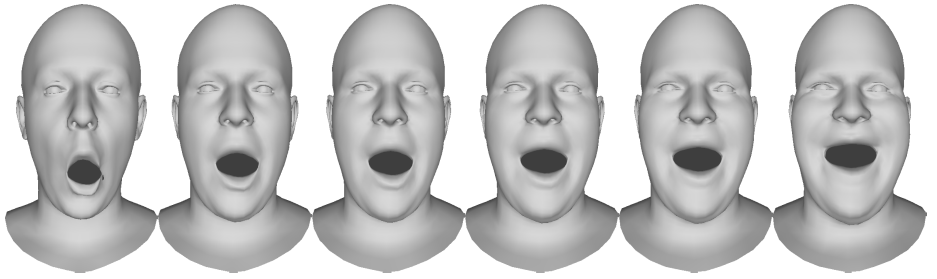}
\caption{Interpolation results between expressions on the COMA dataset.}
\label{fig:results_interpolation}
\end{figure}

\textbf{Interpolation:} To evaluate the interpolation capabilities of our model, we select two significantly different samples, \( x_1 \) and \( x_2 \), from the test set. These samples are encoded into their respective latent representations, $\textbf{z}_1$ and $\textbf{z}_2$, similar to \cite{bouritsas2019_spiral}. Intermediate latent encodings are then generated by interpolating along the line segment connecting $\textbf{z}_1$ and $\textbf{z}_2$, defined as:

\begin{equation}
    \textbf{z} = a \cdot \textbf{z}_1 + (1 - a) \cdot \textbf{z}_2
\end{equation}
where \( a \in [0, 1] \). Fig. \ref{fig:results_interpolation} shows the interpolation results.

\begin{figure}[!t]
\centering
\includegraphics[width=1\linewidth]{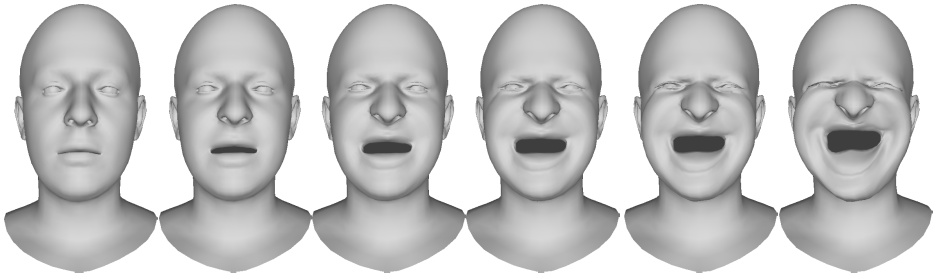}
\caption{Extrapolation results on COMA dataset. Left: neutral expression.}
\label{fig:results_extrapolation}
\end{figure}

\textbf{Extrapolation:} Similarly to interpolation, we evaluated the extrapolation capabilities of our model. We select two significantly different samples and encode them into their respective latent representations, $\textbf{z}_1$ and $\textbf{z}_2$. Extrapolated latent encodings are then generated by extending beyond the line segment connecting $\textbf{z}_1$ and $\textbf{z}_2$, defined as:
\begin{equation}
    \textbf{z} = \textbf{z}_1 + a \cdot (\textbf{z}_2 - \textbf{z}_1)
\end{equation}
where \( a \) can be any real number, including values outside the interval \( [0, 1] \). This allows for the exploration of the latent space beyond the original samples. Fig. \ref{fig:results_extrapolation} shows the extrapolation results.

\begin{figure}[!t]
\centering
\includegraphics[width=1\linewidth]{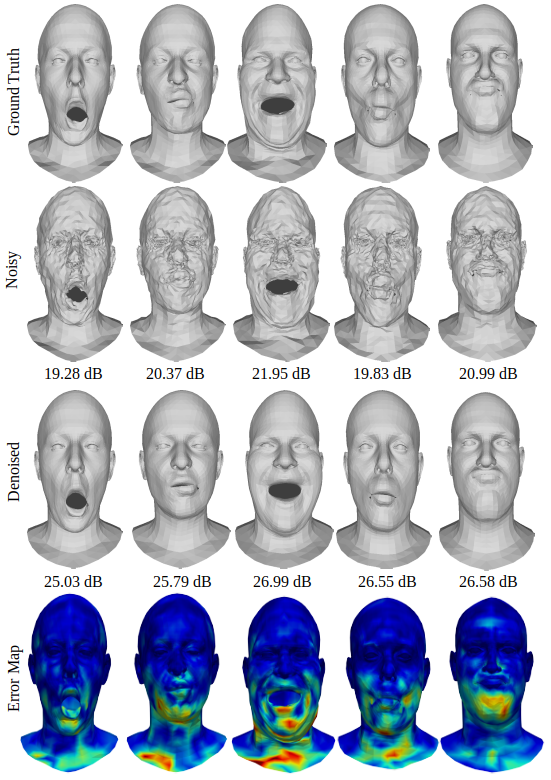}
\caption{
Mesh denoising results for various expressions from the COMA dataset. The PSNR values between the ground truth, noisy, and denoised meshes are displayed below each corresponding sample. The last row illustrates the error maps comparing the ground truth and denoised meshes.}
\label{fig:results_denoising}
\end{figure}

\textbf{Mesh Denoising:} For the second application of our model, we introduced synthetic Gaussian noise to the input mesh data by sampling from a Gaussian distribution with a mean of 0.0 and a standard deviation of $0.001$ to assess its ability to remove noise effectively. 
The denoising results, illustrated in Fig. \ref{fig:results_denoising}, demonstrate that our method successfully removed Gaussian noise while preserving the original geometric details of the mesh. Notably, the model was trained exclusively on noise-free meshes, yet it exhibited robust denoising capabilities across all tested samples.

\section{CONCLUSIONS AND FUTURE WORKS}
In this study, we present 3DGeoMeshNet, a multi-scale GCN for 3D mesh representation and reconstruction. The proposed framework utilizes an autoencoder-decoder architecture comprising complementary local and global networks. The global network is optimized for capturing macro-level mesh characteristics, including overall shape topology, while the local network focuses on micro-level details, such as facial expressions. Through the integration of features from both networks, we achieve a more comprehensive representation by synthesizing information across multiple scales.
Experimental results conducted on the widely utilized COMA human facial dataset demonstrate that our approach yields statistically significant improvements over SOTA methods. The current implementation is constrained to non-textured, monochromatic meshes; a logical extension of this research will involve the incorporation of textured mesh data. Furthermore, while the present approach operates on meshes with fixed vertex counts and topologies, future investigations will focus on generalizing the framework to accommodate meshes with heterogeneous topologies and variable vertex counts, thereby enhancing the robustness and versatility of the proposed methodology.

\bibliographystyle{IEEEtran} 
\bibliography{egbib}        

\end{document}